\documentclass[10pt,letterpaper]{article}
\usepackage{opex3}		

\usepackage{amsmath,amssymb}
\usepackage{graphicx}
\usepackage{cite}

\newtheorem{theorem}{Theorem}[section]

\newtheorem{remark}[theorem]{Remark}

\newcommand{\BE}{\begin{equation}}
\newcommand{\EE}{\end{equation}}

\newcommand{\cO}{\mathcal{O}}

\newcommand{\mR}{\mathbb{R}}
\newcommand{\e}{\mathrm{e}}
\newcommand{\IN}{0}
\newcommand{\uin}{u_\IN}
\newcommand{\rhoin}{\rho_\IN}
\newcommand{\phin}{\phi_\IN}
\newcommand{\vin}{v_\IN}
\newcommand{\eps}{\epsilon}
\newcommand{\sig}{\sigma}

\newcommand{\pt}{\partial}

\newcommand{\bk}[1]{{\langle #1 \rangle}}
\newcommand{\abs}[1]{{\vert {#1} \vert}}

\newcommand{\ovl}[1]{\overline{#1}} 

\DeclareMathOperator{\RE}{Re}
\DeclareMathOperator{\IM}{Im}
\DeclareMathOperator{\rect}{rect}
\DeclareMathOperator{\sinc}{sinc}

\DeclareMathOperator{\sech}{sech}
\DeclareMathOperator{\csch}{csch}

\begin{document}

\title{Signal-noise interaction in nonlinear optical fibers: a hydrodynamic approach}

\author {Luigi Barletti\textsuperscript{1}  and Marco Secondini\textsuperscript{2,*} }

\address{\textsuperscript{1}Dipartimento di Matematica e Informatica ``U.\ Dini'', Universit\`a degli Studi di Firenze, 
Viale Morgagni 67/A, 50134 Firenze, Italy.\\
\textsuperscript{2}TeCIP Institute, Scuola Superiore Sant'Anna, via Moruzzi 1, 56124 Pisa, Italy.}

\email{\textsuperscript{*}marco.secondini@sssup.it} 


\begin{abstract} 
We present a new perturbative approach to the study of signal-noise interactions in nonlinear optical fibers.
The approach is based on the hydrodynamic formulation of the nonlinear Schr\"odinger equation that governs the 
propagation of light in the fiber.
Our method is discussed in general and is developed in more details for some special cases, 
namely the small-dispersion regime, the continuous-wave (CW) signal and the solitonic pulse.
The accuracy of the approach is numerically tested in the CW case.
\end{abstract}


%
\section{Introduction}

In fiber-optic communication systems, optical amplifiers are usually employed to periodically compensate for fiber loss. As a side effect of amplification, amplified spontaneous emission (ASE) noise is added to the optical signal at each amplification stage.	
Both signal and noise propagate through the optical fiber according to the nonlinear Schr\"odinger equation (NLSE), which accounts for attenuation, dispersion, and Kerr nonlinearity. In the linear regime (i.e., at low power), the ASE noise is not affected by propagation through the fiber and can be modeled as additive white Gaussian noise (AWGN) at the end of the optical link. On the other hand, when nonlinear effects are not negligible, signal and noise interact during propagation. This interaction manifests, for instance, in the parametric gain and modulation instability effects \cite{Kar:JOSB1195} and in the emergence of nonlinear phase noise \cite{Gor:OLE1290}. As a result, noise becomes, in general, colored, non-stationary, and non-Gaussian.
The knowledge of the statistics of the noisy signal at the output of the link is extremely important to design efficient optical systems, evaluate their performance, and establish the ultimate channel capacity. For instance, optimum detection (in a maximum likelihood or maximum a-posteriori probability sense) is based on the knowledge of the conditional distribution of the output signal given the input \cite{Marsella:JLT14}. Unfortunately, this distribution is, in general, unknown. Exact statistical models in the absence of fiber dispersion have been studied in \cite{Mec:JLT1994,Ho:2005,Beygi:JLT2011}.
 In the presence of dispersion, approximate models based on perturbation methods have been investigated in \cite{Car:PTL0497,Hol:JLT0302,Ser:JLT0805}.
  Some alternative (non-regular) perturbation methods have been also proposed to describe the noise as a non-additive perturbation of the noise-free solution. For instance, the combined regular-logarithmic perturbation (CRLP) \cite{Secondini09} provides an accurate description of the non-Gaussian distribution through a non-linear combination of Gaussian variables, accounting for both parametric gain and nonlinear phase noise. 
Moreover, in \cite{Marsella:JLT14} it is shown that a polar-Gaussian metric may provide a better performance than a Cartesian-Gaussian metric for maximum likelihood sequence detection. This suggests that a perturbation approach applied to a polar representation (e.g., amplitude and phase) of the signal may be a good alternative to obtain a simple but effective statistical model.
\par
A similar problem arises in the different context of high-power fiber lasers, in which an incoherent or partially-coherent CW light is subject to spectral broadening during propagation through an optical fiber \cite{Manassah1991self,mussot2004spectral,barviau2006spectral,soh2010effect}. Also in this context, the important role of the interaction between fiber dispersion and nonlinearity and the suitability of a polar representation of the signal have been widely recognized.
\par
In this paper we develop a novel approach to the problems outlined above, which is based on the hydrodynamic formulation
of the NLSE.
Such formulation, originally introduced by E.\ Madelung in a quantum mechanical framework \cite{Madelung} and subsequently applied also to optical fibers (e.g, in \cite{kodama1995analytical,Kamchatnov1999,biondini2006whitham,wabnitz2013shallow}), consists of a system of compressible, isothermal,  Euler-like equations with a typical ``quantum pressure'' term, also known as Bohm potential.
In such system the hydrodynamic variables have the physical meaning of power and angular frequency (playing the role
of the fluid density and velocity, respectively).
When the dispersion is small compared to the nonlinearity, the quantum pressure term can be formally neglected and we are left with a much simpler hydrodynamic system which is similar in form to the so called ``shallow-water equations''.
Such a semiclassical approximation has been employed, for instance, in the study of self-similar propagation of high-power pulses in optical fiber amplifiers \cite{anderson1993wave,kruglov2000self,fermann2000self,boscolo2002self}.
It must be noticed, however, that in this way the dynamics loses its dispersive character, which may lead to formation of shocks or even to ill-posedness of the mathematical problem \cite{Levermore}.
Both the full and the approximated hydrodynamic equations are then used to study the propagation of an input signal affected by the ASE noise.
Assuming the mean power $2\sigma^2$ of the injected noise to be small with respect to the typical power of the signal,  we perform a perturbation expansion at first-order  in the small parameter $\sigma$.
The leading order describes the fully deterministic propagation of the signal while the first-order obeys a linear system with stochastic input.
Assuming the input ASE to be described by a band-limited white gaussian process, it turns out that, at first order in $\sigma$, 
the input power and frequency are still described by a joint gaussian process.
Hence, the linearity of the stochastic system implies that the statistics remain gaussian after the propagation, 
which simplifies the computation of the output statistics.
\par
The paper is organized as follows. 
The NLSE is recast in hydrodynamic form in Section \ref{S2}, where also the small-dispersion regime is considered.
In Sec.\ \ref{S3}, the statistics of the hydrodynamic variables are computed up to terms of order $\sigma^2$.
The perturbative expansion is introduced in general in Sec.\ \ref{S4} and is applied to the particular regimes in the subsequent sections \ref{S5} (small-dispersion regime), \ref{S6} (CW signal) and \ref{S7} (solitonic pulse).
In Sec.\ \ref{S8} the perturbative method is used to get equations that govern the evolution of the expected value of the solution. 
Finally, Sec.\ \ref{S9} is devoted to numerical experiments, in the CW case, that are aimed to test the accuracy of the method.
\section{Hydrodynamic formulation of the NLSE}
\label{S2}
We assume, as usual, that the electromagnetic field propagates along the fiber according
to the NLSE \cite{Agrawal}
\BE
\label{NLSE}
  i u_z = -\frac{\beta_2}{2}\,u_{tt}  + \gamma\, \abs{u}^2 u - i\frac{\alpha}{2} u,
\EE
where the subscripts denote derivation with respect to $z$ and $t$.
Here, $z >0$ is a longitudinal coordinate of the fiber axis, $t \in \mR$ is the time coordinate corresponding to a time
frame moving with the signal group velocity, $\beta_2 \in \mR$  is the group velocity dispersion parameter
and $\gamma \geq 0$ and $\alpha \geq 0$ are, respectively, the nonlinear refraction and attenuation parameters. 
The variable $u = u(z,t)$ is the envelope of the modulation of the electromagnetic field along the fiber.
The dispersion parameter $\beta_2$ can be either positive (normal dispersion),
or negative (anomalous dispersion), giving the NLSE, Eq.\ \eqref{NLSE}, a defocusing
or focusing character, respectively.
\par
We apply to Eq.\ \eqref{NLSE} the transformation 
\BE
\label{MadTransf}
  u = \sqrt{\rho}\,\e^{i\phi}, \qquad v = \phi_t,
\EE
where the real quantities  $\rho \geq 0$, $\phi$ and $v$, which are functions of $z$ and $t$, are, respectively, 
power, phase, and angular frequency of the signal $u$.
Substituting Eq.\ \eqref{MadTransf} into Eq.\ \eqref{NLSE}, and writing separately the equations for the real and imaginary
part, we obtain the ``hydrodynamic'' formulation of the NLSE:
\BE
\label{MadEqs}
  \left\{
  \begin{aligned}
  &\rho_z + \beta_2\,(\rho v)_t  = -\alpha\rho,
  \\
  &v_z + \Big(\frac{\beta_2}{2} v^2 + \gamma\rho
   -\beta_2 \frac{(\sqrt{\rho})_{tt}}{2\,\sqrt{\rho}}\Big)_t = 0.
  \end{aligned}
  \right.
\EE
Apart from the exchange of space and time coordinates, the presence of the dissipative term 
$-\alpha\rho$ and, of course, different physical constants, system \eqref{MadEqs} is the hydrodynamic 
form of Schr\"odinger equation (for the nonlinear potential $V = \gamma\rho$) that was introduced by 
E.\ Madelung in 1926 \cite{Madelung}.
They have the form of an Euler system of conservation laws for an isothermal, compressible gas with a 
``quantum pressure'' term $\beta_2Q$, where 
\BE
\label{Qdef}
Q = - \frac{(\sqrt{\rho})_{tt}}{2\,\sqrt{\rho}} 
= \frac{(\rho_t)^2 - 2\rho \rho_{tt}}{8\rho^2}.
\EE
By considering a reference time $T_0$ and power $P_0$ (determined by the duration and mean power of a pulse 
injected in the fiber), it is known that the dimensionless parameter
\BE
\label{eps1}
  \eps = \frac{1}{T_0} \sqrt{\frac{\abs{\beta_2}}{\gamma P_0}},
\EE
measures the relative strength of the dispersion term \cite{Agrawal} 
(it plays the role of a scaled Planck constant for the NLSE \eqref{NLSE}).
By using dimensionless variables, it is possible to show that the error in neglecting the term $\beta_2 Q$ 
in Eq.\ \eqref{MadEqs} is of order $\eps^2$.
Then, for $\eps \ll 1$ we can formally write the approximated system 
\BE
\label{AMad}
  \left\{
  \begin{aligned}
  &\rho_z + \beta_2(\rho v)_t  = -\alpha\rho,
  \\
  &v_z + \Big(\frac{\beta_2}{2}\,v^2 + \gamma\rho \Big)_t = 0.
  \end{aligned}
  \right.
\EE
For rigorous results about the limit $\eps \to 0$ of system \eqref{MadEqs} in the defocusing
case, one can refer to \cite{Levermore}.
\begin{remark}
\rm
Condition $\eps \ll 1$ can be interpreted as a condition on the typical frequency
\BE
\label{eps4}
  \frac{1}{T_0} \ll \sqrt{\frac{\gamma P_0}{\abs{\beta_2}}}.
\EE
which is generally not satisfied in typical optical fiber communication systems.Thus, while the approximated system \eqref{AMad} can be used for specific applications (e.g., high-power fiber lasers or optical amplifiers), we will employ the complete system \eqref{MadEqs} to obtain some results of more general validity, suitable also for optical communications.
\end{remark}
\begin{remark}
\label{rem1}
\rm
Equation \eqref{AMad} with $\alpha = 0$ and $\beta_2>0$
is a hyperbolic system of conservation laws (indeed, it is formally equivalent to the well known 
equations describing shallow waters).
For $\beta_2 <0$ such system is not hyperbolic, which implies that the input-value problem 
is not well posed in general  and leads to unphysical results \cite{Ockendon}. 
This can be clearly seen in the simulations reported in Sec.\ \ref{S9},
where the small-dispersion approximation applied to a fiber with anomalous dispersion predicts
a non-physical, exponential growth.
\end{remark}
%
%
\smallskip
System \eqref{MadEqs}, or its small-dispersion approximation, Eq.\  \eqref{AMad}, will be supplemented with 
input conditions (i.e.\ conditions at $z=0$) of stochastic nature, which will be described in next section.
\section{Stochastic input data}
\label{S3}
We assume that an optical amplifier injects in the fiber, at $z = 0$, an input signal of the form
\BE
\label{inputu}
  u(0,t) = u_\IN(t) + \sig Z(t),
\EE
where $u_\IN(t)$ is the deterministic part of the input signal, and $\sig Z(t)$ is the stochastic part, 
that is the ASE noise.
We assume that $Z(t) = Z_1(t) + i Z_2(t)$ is a gaussian, band-limited, complex white noise
\cite{Papoulis}, i.e.\ a stochastic process such that:
\begin{enumerate}
\item[ i)]
for every fixed $t \in \mR$,  $Z_1(t)$ and $Z_2(t)$ have normal distribution with mean 0 and variance 1
(so that  the variance of the noise is $2\sig^2$), 
and are independent;
\item[ ii)] 
the autocorrelation function is
\BE
\label{autcorrZ}
 \bk{Z_i(t+\tau) Z_j(t)} = \delta_{ij} \frac{1}{\omega_c} \frac{\sin(\omega_c \tau)}{\tau}
 = \delta_{ij} \sinc(\omega_c\tau)
\EE
(where $\bk{\cdot}$ denotes the expectation), corresponding to a power spectral density 
\BE
\label{PSDZ}
 \int_{-\infty}^{+\infty}  \bk{Z_i(t+\tau) Z_j(t)} \e^{-i\omega\tau} d\tau 
 = \delta_{ij} \frac{\pi}{\omega_c} \rect(\omega/2\omega_c),
\EE
where $\omega_c > 0$ is the cutoff frequency. 
\end{enumerate}
Note that $2\sig^2$ is the mean power of the noise and in the following we will assume
\BE
  \sig^2 \ll P_0.
\EE
\subsection{Statistics of the input power and phase}
\label{InputFluid}
The input condition, Eq.\  \eqref{inputu}, has now to be translated into an input condition for the hydrodynamic 
system, Eq.\ \eqref{AMad}. 
This will be done at first order in $\sigma$ (i.e.\ up to terms of order $\sig^2$, that are small compared 
to the typical pulse power $P_0$).
\par
As far as the power is concerned, we can write
$$
  \abs{\uin + \sig Z}^2 = \abs{\uin}^2 + 2\sig\RE(\uin\ovl{Z}) + \cO(\sig^2).
$$
Hence, up to $\cO(\sig^2)$, 
\BE
\label{InRho}
  \abs{u(0,t)}^2 = \rho(0,t) \approx  \rhoin(t) + \sig R(t),
\EE
where, according to Eq.\ \eqref{MadTransf}, the deterministic part of the input has been decomposed as
\BE
\label{fMad}
\uin = \sqrt{\rhoin}\,\e^{i\phin}.
\EE
The stochastic perturbation of the input power,
\BE
\label{Rdef}
  R(t) = 2\RE\left(\uin(t) \ovl{Z}(t)\right),
\EE
is therefore a gaussian process with 0 mean and (by Eq.\ \eqref{autcorrZ}) autocorrelation 
\BE
\label{RAut}
  \bk{R(t) R(s)} 
  = 4\sqrt{\rhoin(t)\rhoin(s)} \,\cos\!\left(\phin(t) - \phin(s)\right)\sinc(\omega_c(t-s)).
\EE
As far as the phase is concerned, we have
$$
  \arg(\uin + \sig Z) = -i \log\left(\frac{\uin+ \sig Z}{\abs{\uin+ \sig Z}} \right)
 = \arg(\uin) - \sig\,\frac{\IM(\uin\ovl{Z})}{\abs{\uin}^2} + \cO(\sig^2).
$$
Since $\phi = \arg(u)$, we have that, up to $\cO(\sig^2)$,
\BE
  \phi(0,t) \approx \phin(t) + \sig\Phi(t),
\EE
where the perturbation 
\BE
\label{PhiDef}
 \Phi(t) = -\frac{\IM\left( \uin(t)\ovl{Z}(t)\right)}{\rhoin(t)}
\EE
is a gaussian process with 0 mean and autocorrelation
\BE
\label{PhiAut}
 \bk{\Phi(t)\Phi(s)} = \frac{\RE\left(\uin(t) \ovl{\uin}(s)\right)}{\rhoin(t) \rhoin(s)} \,\sinc\!\left(\omega_c(t-s) \right).
\EE
%
\subsection{Statistics of the input frequency}
We can now compute the statistics of the input frequency
\BE
\label{InChirp}
  v(0,t) = \frac{d}{dt} \phi(0,t) \approx \vin(t) + \sig V(t),
\EE
where, $\vin(t) = \phin'(t)$ and $V(t) = \Phi'(t)$.
By standard results on differentiated processes \cite{Papoulis} we have that $V(t)$ is a gaussian
process with 0 mean and autocorrelation function
$$
 \bk{V(t)V(s)} = \frac{\pt^2}{\pt t \,\pt s} \bk{\Phi(t)\Phi(s)}.
$$
Then, by using Eq.\ \eqref{PhiAut}, we obtain 
\begin{multline}
\label{ChirpAut}
   \bk{V(t)V(s)} =
    \left\{  \omega_c \left[ \RE\left( \vartheta(t) \ovl{\vartheta'(s)}\right) 
   -  \RE\left(\vartheta'(t)\ovl{\vartheta(s)}\right)\right]  \sinc'\!\left(\omega_c(t-s) \right)
   \right.
 \\[4pt]
 + \left. \RE\left(\vartheta'(t)\ovl{\vartheta'(s)}\right) \, \sinc\!\left(\omega_c(t-s) \right)  
 - \omega_c^2 \, \RE\left(\vartheta(t)\ovl{\vartheta(s)}\right) \, \sinc''\!\left(\omega_c(t-s) \right) \right\},
\end{multline}
where 
$$
  \vartheta := \frac{\uin}{\rhoin} = \frac{\e^{i\phin}}{\sqrt{\rhoin}}
  \qquad \text{and} \qquad 
  \vartheta' =  \frac{\e^{i\phin}}{\sqrt{\rhoin}} 
  \left( i\vin  - \frac{\rho'_0 }{2\rhoin} \right).
$$
\par 
\medskip
We finally compute the cross-correlation between $R$ and $V$.
First of all, from Eqs.\ \eqref{autcorrZ}, \eqref{Rdef} and \eqref{PhiDef}, we immediately obtain 
\BE
\label{RPhiCorr}
  \bk{R(t)\Phi(s)} = 2\IM\left( \uin(t)\ovl{\vartheta(s)}\right) \sinc\!\left(\omega_c(t-s) \right).
\EE
Then, using $\bk{R(t)V(s)} = \frac{\pt}{\pt s}  \bk{R(t)\Phi(s)}$, we have
\BE
\label{RVcorr}
   \bk{R(t)V(s)} =
   2\IM\left(\uin(t) \ovl{\vartheta'(s)}\right) \sinc\!\left(\omega_c(t-s) \right) 
   - 2\omega_c \IM\left(\uin(t)\ovl{\vartheta(s)}\right) \sinc'\!\left(\omega_c(t-s) \right).
\EE
%
%
\section{Perturbative treatment of the stochastic problem}
\label{S4} 
Let us rewrite here the stochastic input-value problem given by Eqs.\ \eqref{MadEqs}, \eqref{InRho} 
and \eqref{InChirp}:
\begin{subequations}
\label{HModel}
\begin{align}
\label{HModeleq}
  &\left\{
  \begin{aligned}
  &\rho_z + \beta_2 (\rho v)_t  = -\alpha\rho,
  \\
  &v_z + \Big(\frac{\beta_2}{2}\, v^2 + \gamma\rho
   -\beta_2\,\frac{(\sqrt{\rho})_{tt}}{2\sqrt{\rho}} \Big)_t = 0.
  \end{aligned}
  \right.
\\[6pt]
\label{HModelcc}
 &\left\{
 \begin{aligned}
 &\rho(0,t) = \rhoin(t) + \sig R(t)
\\
 &v(0,t) =  \vin(t) + \sig V(t)
\end{aligned}
\right.
\end{align}
\end{subequations}
This problem will be treated perturbatively with respect to the small parameter $\sig$;
then we expand the unknowns as
$$
  \rho = \rho^{(0)} + \rho^{(1)} + \cO(\sig^2),
  \qquad
  v = v^{(0)} + v^{(1)} + \cO(\sig^2),
$$
where we assume $\rho^{(1)} \sim v^{(1)} \sim \sigma$.
Substituting into Eq.\ \eqref{HModel} yields, at leading order,
\begin{subequations}
\label{HM0}
\BE
\label{HM0eq}
  \left\{
  \begin{aligned}
  &\rho^{(0)}_z + \beta_2(\rho^{(0)} v^{(0)})_t  = -\alpha\rho^{(0)},
  \\
  &v^{(0)}_z + \Big(\frac{\beta_2}{2}  (v^{(0)})^2 + \gamma\rho^{(0)}
   -\beta_2\,\frac{\big(\sqrt{\rho^{(0)}} \big)_{tt}}{2\sqrt{\rho^{(0)}}} \Big)_t = 0,
  \end{aligned}
  \right.
\EE
with input conditions
\BE
\label{HM0ic}
 \left\{
 \begin{aligned}
 &\rho^{(0)}(0,t) = \rhoin(t)
\\
 &v^{(0)}(0,t) =  \vin(t)
\end{aligned}
\right.
\EE
\end{subequations}
(which, of course, means that at leading order we are left with the deterministic problem)
and, at first order,
\begin{subequations}
\label{HM1}
\begin{align}
\label{HM1eq}
   &\left\{
  \begin{aligned}
  &\rho^{(1)}_z + \beta_2 {\left(  v^{(0)}  \rho^{(1)}  + \rho^{(0)}v^{(1)}   \right)}_t
  = -\alpha\rho^{(1)},
  \\
  &v^{(1)}_z + {\left[ \beta_2\, v^{(0)} v^{(1)} + \gamma \rho^{(1)}
  +\beta_2\, Q^{(1)} \right]}_t = 0,
  \end{aligned}
  \right.
\\[6pt]
\label{HM1ic}
 &\left\{
 \begin{aligned}
 &\rho^{(1)}(0,t) = \sig R(t)
\\
 &v^{(1)}(0,t) =  \sig V(t)
\end{aligned}
\right.
\end{align}
\end{subequations}
where  $Q^{(1)}$ is the first-order approximation of $Q$ (see Eq.\ \eqref{Qdef}), given by
\BE
\label{Q1def}
  Q^{(1)}= 
  \frac{\rho^{(1)}}{4\rho^{(0)}} \Big(\frac{\rho^{(0)}_t}{\rho^{(0)}}\Big)_t  
  -  \frac{1}{4} \Big(\frac{\rho^{(1)}_t}{\rho^{(0)}}\Big)_t   
= \frac{\rho^{(1)}}{4\rho^{(0)}} \big( \log \rho^{(0)} \big)_{tt}  
-  \frac{1}{4}\Big(\frac{\rho^{(1)}_t}{\rho^{(0)}}\Big)_t 
\EE
(linear in $\rho^{(1)}$).
Note that Eq.\ \eqref{HM0} is a nonlinear deterministic problem (equivalent to the NLSE) while, once 
Eq.\ \eqref{HM0} has been solved, Eq.\ \eqref{HM1} is a linear stochastic problem.
\begin{remark}
\rm
The stochastic problem introduced in this section refers to the case of signal and ASE noise injected at the input of a single span of fiber. The analysis can be easily extended to a multi-span optical system by following the same approach employed in \cite{Secondini09}. In this case, at each amplification stage, an independent perturbation of the order of $\sigma$ (due to the ASE noise generated by the inline amplifier), with statistics as in Eqs.\ \eqref{RAut}, \eqref{ChirpAut} and \eqref{RVcorr}, is added to the (already noisy) Madelung variables $\rho$ and $v$, which are then propagated through the following span of fiber according to Eq.\ \eqref{HModeleq}.
\end{remark}
\section{The small-dispersion approximation}
\label{S5} 
Let us consider the small-dispersion approximation ($Q \approx 0$) of the input-value problems
\eqref{HM0} and \eqref{HM1}.
The leading order equation, Eq.\ \eqref{HM0eq}, becomes
\BE
\label{HM0A}
  \left\{
  \begin{aligned}
  &\rho^{(0)}_z + \beta_2(\rho^{(0)} v^{(0)})_t  = -\alpha \rho^{(0)},
  \\
  &v^{(0)}_z + \Big( \frac{\beta_2}{2} (v^{(0)})^2 + \gamma\rho^{(0)} \Big)_t = 0,
  \end{aligned}
  \right.
\EE
that is the ``shallow water'' system \eqref{AMad}.
As already mentioned in Remark \ref{rem1}, according to the standard theory of first-order systems
\cite{Ockendon,Whitham}, we have that the system is hyperbolic if and only if $\beta_2 > 0$. 
In this case the characteristic velocities are
\BE
\label{charvel}
  c_\pm = \beta_2 v^{(0)} \pm \sqrt{\beta_2 \gamma \rho^{(0)}},
\EE
and the characteristic curves are defined by the equations
\BE
\label{charcur}
  \frac{dt}{dz} = c_\pm =\beta_2 v^{(0)} \pm \sqrt{\beta_2 \gamma \rho^{(0)}}.
\EE
Moreover, we obtain the equations
\BE
\label{Riemann1}
  \frac{d}{dz}  \left( 2 \sqrt{\rho^{(0)}} \pm  \sqrt{\frac{\beta_2}{\gamma}}\,  v^{(0)} \right) 
  = -\alpha \sqrt{\rho^{(0)}},
\qquad 
  \text{on} \quad \frac{dt}{dz} = c_\pm,
\EE
for the Riemann variables $2\sqrt{\rho^{(0)}} \pm \sqrt{\frac{\beta_2}{\gamma}}\, v^{(0)}$.
For $a = 0$, the Riemann variables are constants along the characteristic lines, in which case they are 
Riemann invariants for the problem.
The condition $\alpha = 0$ can always be assumed on a short fiber stretch, provided that
the nonlinear coefficient $\gamma$ is substituted with an effective coefficient 
$\gamma_\mathrm{eff}$  \cite{Agrawal,Secondini09}.
\begin{remark}
\rm
Although the characteristic equations, Eq.\ \eqref{Riemann1},
do not lead to the analytic solution of Eq.\ \eqref{HM0A} (because they hold
separately on the two characteristic curves, which depend on the solution itself), nevertheless the 
form Eq.\ \eqref{Riemann1} allows a very efficient numerical treatment of the problem
(see e.g.\ \cite{Whitham}).
\end{remark}
Assume now that $(\rho^{(0)},v^{(0)})$ is the solution to problem \eqref{HM0A}, and 
consider the small-dispersion approximation of the first-order equation \eqref{HM1eq}:
\BE
\label{HM1A}
  \left\{
  \begin{aligned}
  &\rho^{(1)}_z + \beta_2 \big(  v^{(0)}  \rho^{(1)}  + \rho^{(0)}v^{(1)}\big)_t
  = -\alpha\rho^{(1)},
  \\
  &v^{(1)}_z + \big( \beta_2\, v^{(0)} v^{(1)} + \gamma\rho^{(1)} \big)_t = 0,
  \end{aligned}
  \right.
\EE
This is a linear, non-autonomous, problem for the unknowns $(\rho^{(1)},v^{(1)})$.
For $\beta_2 > 0$, its characteristic curves are fixed, and are the same as those of problem \eqref{HM0A}, 
i.e.\ they are still given by Eq.\ \eqref{charcur}.
The characteristic equations read as follows:
\BE
\label{RE} 
   \frac{d}{dz}\,r_\pm +  \beta_2 v^{(0)}_t r_\pm 
  + \left(\beta_2 \rho^{(0)}_t \mp \frac{d}{dz} \sqrt{\frac{\beta_2\rho^{(0)}}{\gamma}}  \right) v^{(1)} 
  =   -\alpha \rho^{(1)}  ,
 \quad \text{on} \ \frac{dt}{dz} = c_\pm,
\EE
where $r_\pm = \rho^{(1)} \pm   \sqrt{\frac{\beta_2\rho^{(0)}}{\gamma}} \, v^{(1)}$
are the Riemann variables for the first-order problem.
\section{The continuous-wave case}
\label{S6}
The simplest case in which problems \eqref{HM0} and \eqref{HM1} can be explicitly solved is that
of constant deterministic input data $\rhoin$ and $\vin$.
We shall also assume $\alpha = 0$.
Under these conditions, problem \eqref{HM0} admits the constant solution
\BE
\label{CW}
  \rho^{(0)}(z,t) \equiv \rhoin, \qquad  v^{(0)}(z,t) \equiv \vin.
\EE
Such solution is physically meaningful and corresponds to a continuous-wave (CW) solution of the 
NLSE \eqref{NLSE} (with $\alpha=0$) \cite{Secondini09}
\BE
\label{CWsol0}
  u(z,t) = 
    \e^{-i\big(\beta_2 \frac{\vin^2}{2} + \gamma\rhoin\big)z}\, \e^{i \vin t}\,\sqrt{\rhoin}
\EE
(up to a constant phase).
The first-order equation, Eq.\  \eqref{HM1eq}, becomes
\BE
  \left\{
  \begin{aligned}
  &\rho^{(1)}_z + \beta_2 v_0 \rho^{(1)}_t + \beta_2 \rho_0 v^{(1)}_t = 0,
\\
  &v^{(1)}_z + \beta_2 v_0 v^{(1)}_t + \Big( \gamma\rho^{(1)} 
  -\frac{\beta_2}{4\rho_0}  \, \rho^{(1)}_{tt}\Big)_t = 0,
  \end{aligned}
  \right.
\EE
At this point we can notice that we can get rid of the terms containing $v_0$ by means of the change of variable 
$t \mapsto t - \beta_2v_0z$ (indeed, $v_0$ just corresponds to a global frequency shift), which leads to
\BE
\label{CRLP2}
  \left\{
  \begin{aligned}
  &\rho^{(1)}_z + \beta_2\rho_0 v^{(1)}_t = 0  
\\
  &v^{(1)}_z + \Big( \gamma\rho^{(1)}  -\frac{\beta_2}{4\rho_0}  \, \rho^{(1)}_{tt} \Big)_t = 0.
  \end{aligned}
  \right.
\EE
In order to make a comparison with the CRLP method  \cite{Secondini09}, we return to the the phase variable
$\phi_t = v$ for which we easily obtain
\begin{subequations}
\label{CRLPsys}
\BE
\label{CRLP3}
  \left\{
  \begin{aligned}
  &\rho^{(1)}_z + \beta_2\rho_0 \,\phi^{(1)}_{tt} = 0 ,
\\
  &\phi^{(1)}_z + \gamma\rho^{(1)} - \frac{\beta_2}{4\rho_0}  \, \rho^{(1)}_{tt} = 0,
  \end{aligned}
  \right.
\EE
with the stochastic input conditions
\BE
\label{CRLPic}
 \left\{
 \begin{aligned}
 &\rho^{(1)}(0,t) = \sig R(t),
\\
 &\phi^{(1)}(0,t) =  \sig\Phi(t).
\end{aligned}
\right.
\EE
\end{subequations}
Note that an arbitrary function of $z$ was set to 0 in Eq.\ \eqref{CRLP3};
this can be justified by looking at the Madelung equation for the phase,
$$
  \phi_z +\frac{\beta_2}{2} (\phi_t)^2 + \gamma\rho + \beta_2 Q  = 0,
$$
of whom Eq.\ \eqref{CRLP3} is the linearization. 
The comparison of Eqs.\ \eqref{CRLPsys} with the CRLP model is not straightforward since, in the latter, a three-variable perturbation expansion is used to represents the optical signal. However, as shown in \cite{Secondini09}, the CRLP expansion can be reduced to a simpler amplitude and phase representation by setting the quadrature component equal to zero. This is a reasonable approximation when nonlinearity dominates over dispersion. In this case, the equation governing the propagation of the CRLP amplitude and phase variables is analogous to Eq.\ \eqref{CRLP3}, but for obvious differences (a factor of two dividing and multiplying the right-hand side of the first and second equation, respectively) due to the fact that  Eq.\ \eqref{CRLP3} refers to a power and phase expansion.
The Fourier transform of  Eq.\ \eqref{CRLP3} is
\BE
\label{CRLP4}
  \left\{
  \begin{aligned}
  &\hat \rho^{(1)}_z - \beta_2\rho_0 \omega^2 \hat \phi^{(1)} = 0,
\\
  &\hat \phi^{(1)}_z + \gamma\hat \rho^{(1)} 
  + \frac{\beta_2}{4\rho_0} \omega^2 \hat \rho^{(1)}= 0,
  \end{aligned}
  \right.
\EE
and the associated transfer matrix $T(z,\omega)$, such that
\BE
\label{CWsol}
\begin{pmatrix}
 \hat\rho^{(1)} \\ \hat\phi^{(1)} 
\end{pmatrix} 
(z,\omega) = T(z,\omega) 
\begin{pmatrix}
 \sig \hat R \\ \sig \hat \Phi  
\end{pmatrix} 
(\omega),
\EE
can be explicitly computed and reads as follows:
\BE
\label{Tmatrix}
 T(z,\omega) = 
 \begin{pmatrix} 
 \cos(z\sqrt{k})
  &
  \sqrt{\frac{4\beta_2\rho_0^2\omega^2}{\beta_2\omega^2 + 4 \gamma\rho_0}}\,  \sin(z\sqrt{k})
\\[4pt]
- \sqrt{\frac{\beta_2\omega^2 + 4 \gamma\rho_0}{4\beta_2\rho_0^2\omega^2}}\, \sin(z\sqrt{k})
& 
\cos(z\sqrt{k}) 
\end{pmatrix}
\EE
where
\BE
\label{kdef}
k = k(\omega) = \beta_2 \omega^2  \left(\frac{ \beta_2\omega^2}{4} + \gamma\rho_0 \right) .
\EE
This expression holds for $\beta_2 >0$, or $\beta_2 < 0$ and  $\omega^2 > \frac{4\gamma\rho_0}{\abs{\beta_2}}$.
The expression of $T(z,\omega)$ for $\beta_2<0$ and $\omega^2 < \frac{4\gamma\rho_0}{\abs{\beta_2}}$
is easily obtained by using 
$\cos(ix) = \cosh(x)$ and $\sin(ix) = i\sinh(x)$ ($x$ real).
\par
In the  small-dispersion approximation, Eq.\ \eqref{CRLP3} drastically simplifies into
\BE
\label{CWA}
  \left\{
  \begin{aligned}
  &\rho^{(1)}_z + \beta_2\rho_0\,\phi^{(1)}_{tt} = 0 ,
\\
  &\phi^{(1)}_z + \gamma\rho^{(1)}  = 0,
  \end{aligned}
  \right.
\EE
and the corresponding transfer matrix, for $\beta_2 > 0$, becomes 
\BE
\label{TmatrixA}
 T(z,\omega) = 
 \begin{pmatrix} 
 \cos(z\omega\sqrt{\beta_2\rho_0\gamma}) 
 & 
 \sqrt{\frac{\beta_2\rho_0}{\gamma}}\,\omega\sin(z\omega\sqrt{\beta_2\rho_0\gamma})
\\[4pt]
 -\sqrt{\frac{\gamma}{\beta_2\rho_0}}\,\frac{1}{\omega}\sin(z\omega\sqrt{\beta_2\rho_0\gamma})
 &  
 \cos (z\omega\sqrt{\beta_2\rho_0\gamma})
 \end{pmatrix},
\EE
while the expression for $\beta_2 < 0$ can be obtained as explained above.
%
%
\subsection{PSD matrix for the continuous wave}
The transfer matrix $T(z,\omega)$ allows to obtain the power spectral density (PSD) matrix
\BE
\label{Gdef}
G(z,\omega) =  \int_{-\infty}^{+\infty}   
\begin{pmatrix} 
\bk{\rho^{(1)}(z,t) \rho^{(1)}(z,t+\tau)} & \bk{\rho^{(1)}(z,t) \phi^{(1)}(z,t+\tau)} 
\\
\bk{\phi^{(1)}(z,t) \rho^{(1)}(z,t+\tau)} & \bk{\phi^{(1)}(z,t) \phi^{(1)}(z,t+\tau)}
\end{pmatrix}
\e^{-i\omega\tau} d\tau, 
\EE
from its input value $G(0,\omega)$, through 
\BE
\label{Gatz}
  G(z,\omega) = T(z,\omega) G(0,\omega) T^*(z,\omega),
\EE
where $T_z^*(\omega)$ denotes the adjoint matrix (which is just the transpose, in this case).
The input PSD matrix 
\BE
G(0,\omega) =  \sig^2 \int_{-\infty}^{+\infty}   
\begin{pmatrix} 
\bk{R(t) R(t+\tau)} & \bk{R(t) \Phi(t+\tau)} 
\\
\bk{\Phi(t) R(t+\tau)}  & \bk{\Phi(t) \Phi(t+\tau)}
\end{pmatrix}
\e^{-i\omega\tau} d\tau, 
\EE
can be easily obtained from Eqs.\ \eqref{RAut}, \eqref{PhiAut} and \eqref{RPhiCorr}, with $\uin = \sqrt{\rho_0}$,  
$\phi_0 = 0$ and $v_0 = 0$,  and reads as follows:
\BE
\label{inputG}
  G(0,\omega) = \frac{\pi\sig^2}{\omega_c}
  \begin{pmatrix}
  4\rhoin & 0 
  \\ 
 0  &  \frac{1}{\rhoin}
 \end{pmatrix}
  \rect\left(\frac{\omega}{2\omega_c} \right).
\EE
Of course, in the ``white noise limit'' $\omega_c \to \infty$ the $\rect$ function tends to 1. 
\section{Perturbation of a soliton}
\label{S7} 
Let us now consider the one-parameter family of solitonic solutions of the 
focusing NLSE (i.e.\ Eq.\  \eqref{NLSE},  with $\beta_2 < 0$ and $\alpha  = 0$):
\BE
\label{soliton}
  u(z,t) = \sqrt{\frac{\abs{\beta_2}}{\gamma}}\, \eta \sech(\eta t)\, \e^{i\beta_2 \eta^2 z/2},
\EE
where $\eta \geq 0$ is a parameter that determines the soliton width.
A deterministic signal of the form \eqref{soliton} is a $z$-independent solution of 
Eq.\ \eqref{HM0eq} with
\BE
\label{soliton0}
  \rho^{(0)} = \frac{\abs{\beta_2}}{\gamma}\,\eta^2  \sech^2(\eta t),
  \qquad
  \phi^{(0)} = \frac{\beta_2\eta^2z}{2},
  \qquad
  v^{(0)} = 0.
\EE
Substituting $\rho^{(0)}$ and $v^{(0)}$ given by Eq.\ \eqref{soliton0} in Eq.\ \eqref{HM1eq} we obtain
an equation for the perturbation
which, in the power-phase representation, takes the reads as follows:
\BE
\label{HM1solitonPP}
  \left\{
  \begin{aligned}
  &\rho^{(1)}_z -\frac{\beta_2^2 \eta^2}{\gamma}  \big( \sech^2(\eta t)\, \phi^{(1)}_t \big)_t 
  = 0,
  \\[6pt]
  &\phi^{(1)}_z
   + \frac{3}{2}\gamma\rho^{(1)}  + \frac{\gamma}{4\eta^2}  \Big(\frac{\rho^{(1)}_t}{ \sech^2(\eta t) } 
   \Big)_t = 0.
  \end{aligned}
  \right.
\EE
Analogous results are obtained for the dark solitons in the defocusing case ($\beta_2 >0$):
\BE
 \label{darksoliton}
  u(z,t) = \sqrt{\frac{\beta_2}{\gamma}}\, \eta \tanh(\eta t)\, \e^{-i\beta_2 \eta^2 z},
\EE
corresponding to
\BE
\label{soliton1}
  \rho^{(0)} = \frac{\beta_2}{\gamma}\,\eta^2  \tanh^2(\eta t),
  \quad
  \phi^{(0)} = \pi\delta(t) -\beta_2\eta^2z,
  \quad
  v^{(0)} = \pi \delta'(t).
\EE
Substituting $\rho^{(0)}$ and $v^{(0)}$ given by  Eq.\ \eqref{soliton0} in  Eq.\ \eqref{HM1eq}, and disregarding 
irrelevant terms coming from the delta function, we obtain the following equation for the perturbation, 
that we write directly in the power-phase representation:
\BE
\label{HM1blackPP}
  \left\{
  \begin{aligned}
  &\rho^{(1)}_z + \frac{\beta_2^2 \eta^2}{\gamma}  \big( \tanh^2(\eta t)\, \phi^{(1)}_t \big)_t 
  = 0,
  \\[6pt]
  &\phi^{(1)}_z
   + \Big(1 - \frac{\csch^2(\eta t)\sech^2(\eta t)}{2\tanh^2(\eta t)}\Big) \gamma\rho^{(1)}  
   - \frac{\gamma}{4\eta^2}  \Big(\frac{\rho^{(1)}_t}{ \tanh^2(\eta t) } \Big)_t 
  = 0.
  \end{aligned}
  \right.
\EE
\par
\bigskip
It may be interesting to look at an approximation up to terms of order $\sigma\eps^2$. 
This amounts to keeping the leading-order problem, Eqs.\ \eqref{HM0}, unchanged and neglecting
the term containing $Q^{(1)}$  (which is of order $\sigma\eps^2$) in the perturbed problem, Eqs.\ \eqref{HM1}.
In the solitonic case, this means that the leading-order solution are still given by Eqs.\ \eqref{soliton0} 
and \eqref{soliton}, but  the second equation in both systems \eqref{HM1solitonPP} and 
\eqref{HM1blackPP} is replaced by $\phi^{(1)}_z +  \gamma\rho^{(1)} = 0$.
Hence, in the focusing case we obtain the second order equation
\BE
\label{WWE}
  \rho^{(1)}_{zz} + \beta_2^2\eta^2  \big( \sech^2(\eta t)\, \rho^{(1)}_t \big)_t =0,
\EE
which is of elliptic type.
In the defocusing case, instead, we obtain the wave equation
\BE
\label{RWE}
  \rho^{(1)}_{zz} - \beta_2^2\eta^2  \big( \tanh^2(\eta t)\, \rho^{(1)}_t \big)_t =0.
\EE
The elliptic character of Eq.\ \eqref{WWE} is another manifestation of the bad behaviour of the small dispersion 
approximation in the focusing case, since it corresponds to the non-hyperbolic character of the first-order
system for $(\rho^{(1)}, \phi^{(1)})$ or $(\rho^{(1)}, v^{(1)})$ (and see Remark \ref{rem1}). 
\section{Equations for the expected value}
\label{S8} 
Let us finally consider the problem of obtaining equations for the expected values $\bk{\rho}$ and 
$\bk{v}$.
Since $\bk{R} = \bk{V} = 0$, taking the expectation of both sides of the first-order problem, Eq.\ \eqref{HM1}, yields
a linear system in $\bk{\rho^{(1)}}$ and $\bk{v^{(1)}}$ with vanishing input data, which has 
the unique solution $\bk{\rho^{(1)}} = \bk{v^{(1)}} = 0$.
\par
Then, in order to see the effects of the noise on $\bk{\rho}$ and $\bk{v}$, we have to look at the next order 
in $\sig$:
$$
  \bk{\rho} = \rho^{(0)} +\bk{\rho^{(2)}} + \cO(\sig^3),
  \qquad
   \bk{v} = v^{(0)} + \bk{v^{(2)}} + \cO(\sig^3).
$$
In fact, by expanding the input Madelung variables $\rho(0,t)$ and $v(0,t)$,
we see that the second-order system has a non-trivial input:
\BE
\label{icmean}
  \bk{\rho^{(2)}}(0,t) = 2\sig^2, \qquad \bk{v^{(2)}}(0,t) = 0,
\EE
(note that $\rho^{(2)}(0,t)/\sig^2 = \abs{Z(t)}^2$ has a $\chi^2(2)$ distribution).
The second-order equation for $(\rho,v)$ is easily obtained from Eq.\ \eqref{HModeleq}
and reads as follows:
\BE
\label{HM2}
  \left\{
  \begin{aligned}
  &\rho^{(2)}_z +\beta_2 \big(  v^{(0)}  \rho^{(2)}  + \rho^{(1)}v^{(1)} +  \rho^{(0)}v^{(2)}  \big)_t
  = -\alpha \rho^{(2)},
  \\
  &v^{(2)}_z + \Big[ \beta_2 v^{(0)} v^{(2)} +\frac{\beta_2}{2}v^{(1)} v^{(1)}  
  + \gamma \rho^{(2)} + \beta_2 Q^{(2)}  \Big]_t = 0,
  \end{aligned}
  \right.
\EE
where $\beta_2Q^{(2)}$ is the second-order approximation of $Q$, given by
\BE
\label{Q2def}
  Q^{(2)} = 
  \frac{2\rho^{(0)}\rho^{(2)}-(\rho^{(1)})^2}{8(\rho^{(0)})^2} \Big(\frac{\rho^{(0)}_t}{\rho^{(0)}} \Big)_t
  -\frac{1}{4}\Big(\frac{\rho^{(2)}_t}{\rho^{(0)}} \Big)_t
  + \frac{1}{8\rho^{(0)}}  \Big(\frac{(\rho^{(1)})^2}{\rho^{(0)}} \Big)_{tt}.
\EE
Taking the expectation of Eq.\ \eqref{HM2}, and recalling that $\rho^{(0)}$ is a deterministic quantity, 
we obtain 
\BE
\label{meanEq}
\left\{
\begin{aligned}
   &\bk{\rho^{(2)}}_z + \beta_2 \big( v^{(0)}\bk{\rho^{(2)}} +  \bk{\rho^{(1)} v^{(1)}}
   + \rho^{(0)} \bk{v^{(2)}}\big)_t  = -\alpha \bk{\rho^{(2)}},
   \\[4pt]
   &\bk{v^{(2)}}_z + \Big[ \beta_2v^{(0)} \bk{v^{(2)}} + \frac{\beta_2}{2} \bk{v^{(1)} v^{(1)}} 
   + \gamma\bk{\rho^{(2)}} + \beta_2 \bk{Q^{(2)}} \big]_t = 0,
\end{aligned}
\right.
\EE
which has to be supplemented with the input conditions \eqref{icmean}.
Note that Eq.\ \eqref{meanEq} is a linear equation for the unknown $\bk{\rho^{(2)}}$ which depends on the
covariances between $\rho^{(1)}$ and $v^{(1)}$.
\par
\medskip
This is an intricate problem in general, but it simplifies in particular situations.
For example, let us consider the CW case, discussed in Sec.\ \ref{S6}.
In this case  $\rho^{(0)} = \rho_0$ and  $v^{(0)} = v_0$ are two constants 
(and we can assume $v_0 = 0$ without loss of generality). 
Moreover, $\rho^{(1)}$ and $v^{(1)}$ are given by Eq.\ \eqref{CRLPsys} and we have 
$\bk{\rho^{(1)} \rho^{(1)}}_t = \bk{\rho^{(1)} v^{(1)}}_t =  \bk{v^{(1)} v^{(1)}}_t = 0$,
since $(\rho^{(1)}, v^{(1)})$ is a stationary process in the CW case.
Then, from Eqs.\ \eqref{meanEq} and \eqref{Q2def}, we obtain (in the power-phase
representation)
\BE
\label{meanCW2}
  \left\{
  \begin{aligned}
  &\bk{\rho^{(2)}}_z + \beta_2 \rho_0\,\bk{\phi^{(2)}}_{tt} = 0,  
\\[4pt]
  &\bk{\phi^{(2)}}_z +  \gamma \bk{\rho^{(2)}} 
  -\frac{\beta_2}{4\rho_0}\,\bk{\rho^{(2)}}_{tt}
  = 0.
  \end{aligned}
  \right.
\EE
\par
Hence, the equation for $\bk{\rho^{(2)}}$ and $\bk{\phi^{(2)}}$ in the CW case has the same form as
Eq.\ \eqref{CRLP3}.
Supplemented with the input conditions \eqref{icmean},
it has the $t$-independent solution 
\BE
\label{meansol}
  \bk{\rho^{(2)}}(z,t) = 2\sig^2, \qquad \bk{\phi^{(2)}}(0,t) = -2\gamma \sig^2 z
\EE
(up to an arbitrary phase), which simply represents the additional power due to the presence of noise and the corresponding nonlinear phase rotation due to the Kerr effect.
\section{Numerical results}
\label{S9}
\begin{figure}[htb]
\centering
\includegraphics[width=0.5\textwidth]{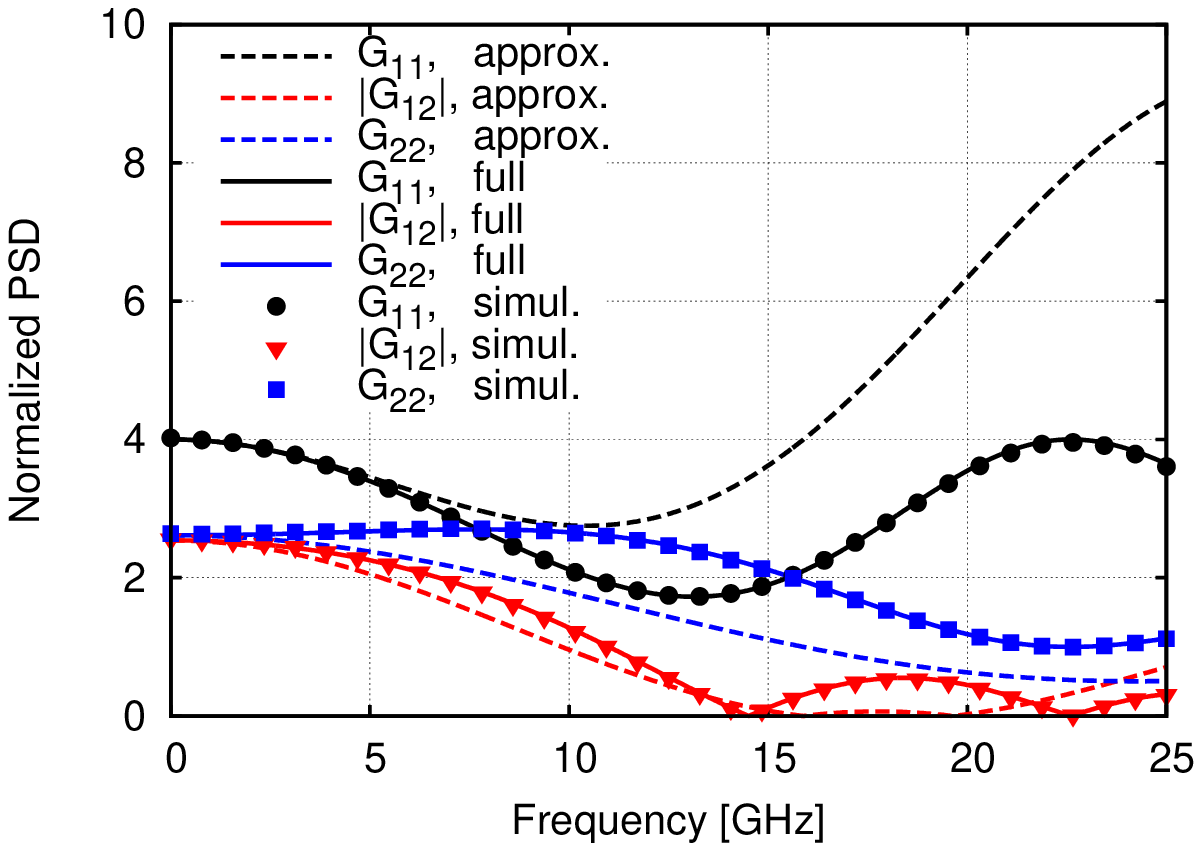}\includegraphics[width=0.5\textwidth]{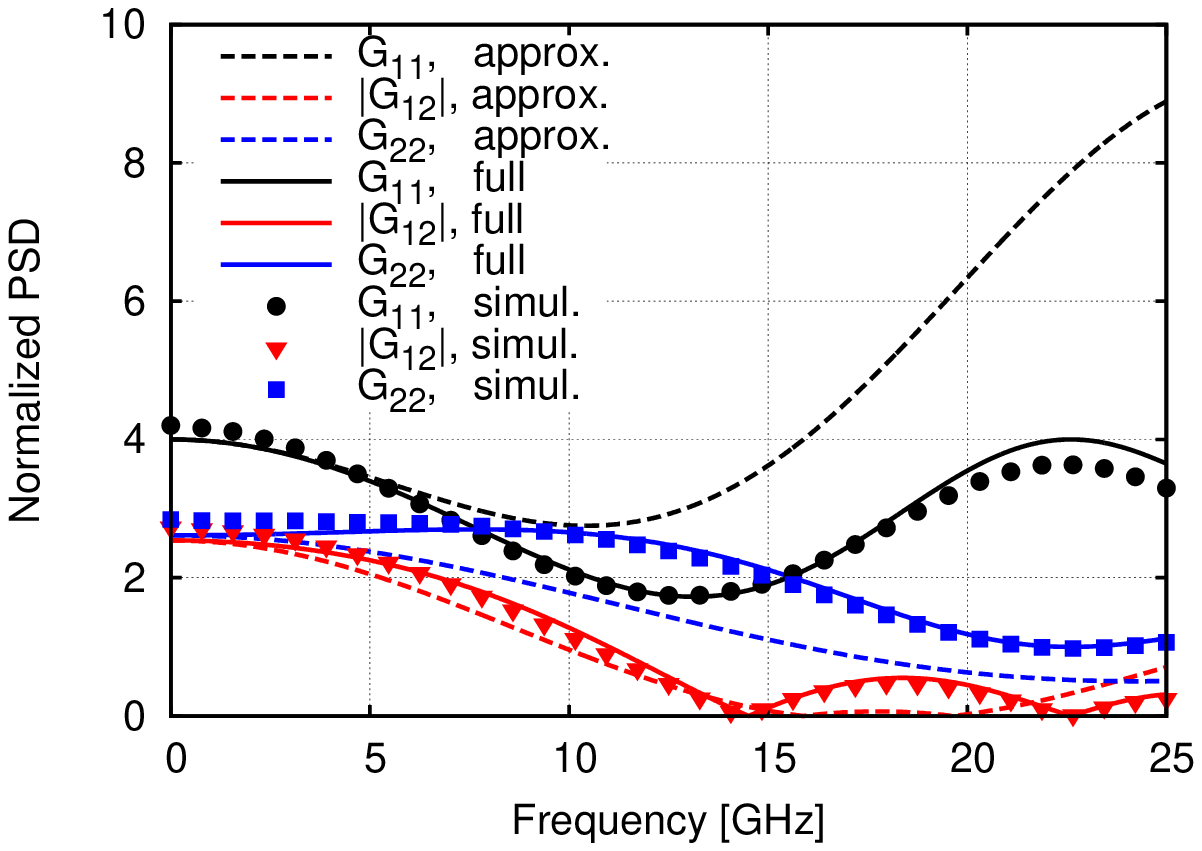} 
\caption{\label{fig:PSD_-4}Normalized PSD matrix at the output of the non-zero dispersion fiber (normal dispersion $\beta_2=5.1\,\mathrm{ps^2/km}$) for an SNR of (a) 20~dB and (b) 10~dB.}
\end{figure}

The accuracy of the proposed approach is tested in the CW case, for which exact analytical solutions for the PSD and pdf of the received signal have been derived. We consider the propagation of a CW signal affected by band-limited AWGN noise through 50 km of lossless fiber. As a first test, we consider a non-zero dispersion-shifted fiber with GVD parameter $\beta_2=5.1\,\mathrm{ps^2/km}$ (normal dispersion)
and nonlinear coefficient $\gamma=1.27\,\mathrm{W^{-1}km^{-1}}$, a signal power of 10~mW, and a noise bandwidth of 50~GHz. The PSD matrix of the output power and phase variables is evaluated analytically from Eq.\ \eqref{Gatz}, by using the transfer matrix either from the full first-order model, Eq.\  \eqref{Tmatrix}, or from the small-dispersion approximation, Eq.\ \eqref{TmatrixA}. 
For comparison, the same PSD matrix is evaluated by means of numerical simulations by employing the split-step Fourier method \cite{Agrawal} to propagate the noisy signal through the fiber, and the averaged periodogram to estimate the PSD.
Figures~\ref{fig:PSD_-4}(a) and \ref{fig:PSD_-4}(b) show the results obtained for an SNR of 20 and 10~dB, respectively. 
In both cases, the significant deviation of the output PSD matrix from the input one, Eq.\  \eqref{inputG} (which is diagonal and frequency-independent within the noise bandwidth), testifies the presence of a significant signal-noise interaction. At 20~dB of SNR, there is a perfect agreement between the full model and numerical simulations, while the approximated model, as expected, is accurate only at low frequency.
On the other hand, at 10~dB of SNR, a small discrepancy between numerical simulations and the full model can be observed. This is due to the neglected higher-order terms in $\sigma$ which, at 10~dB of SNR, become relevant. 
\par
\begin{figure}[htb]
\centering
\includegraphics[width=0.5\textwidth]{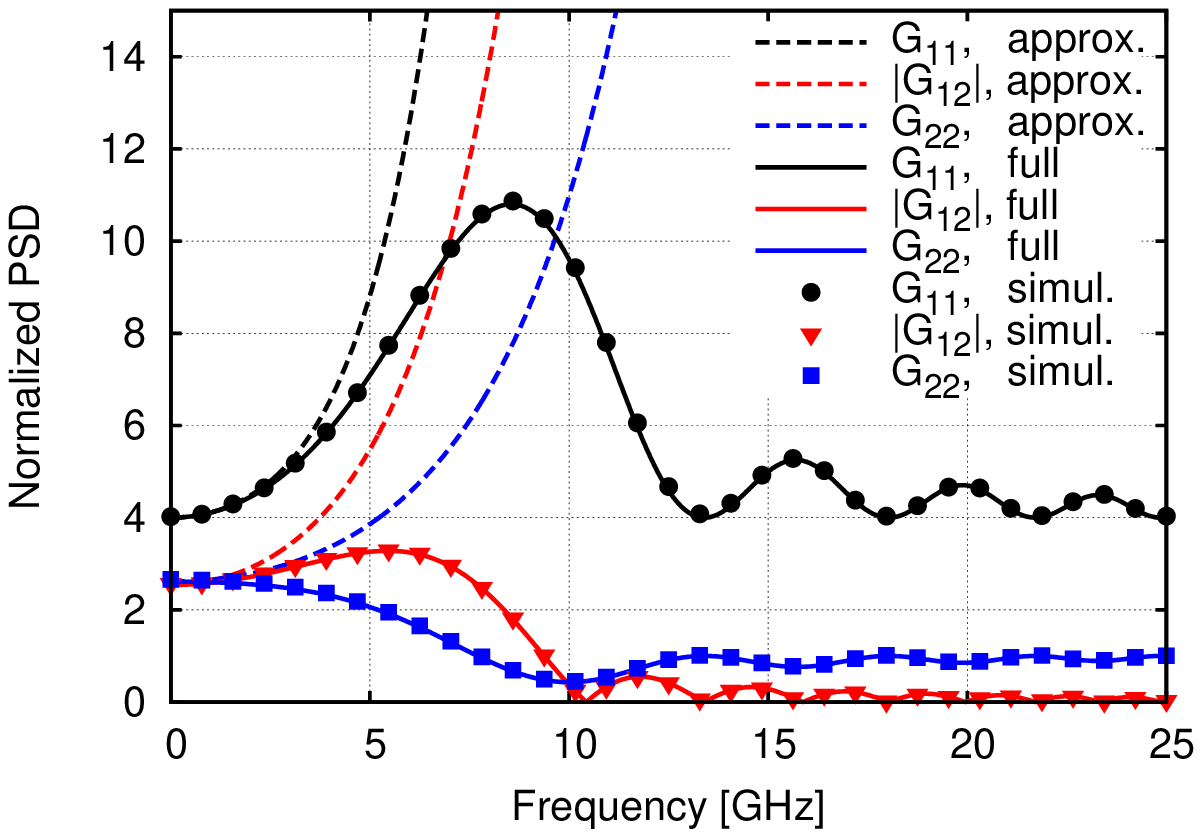}\includegraphics[width=0.5\textwidth]{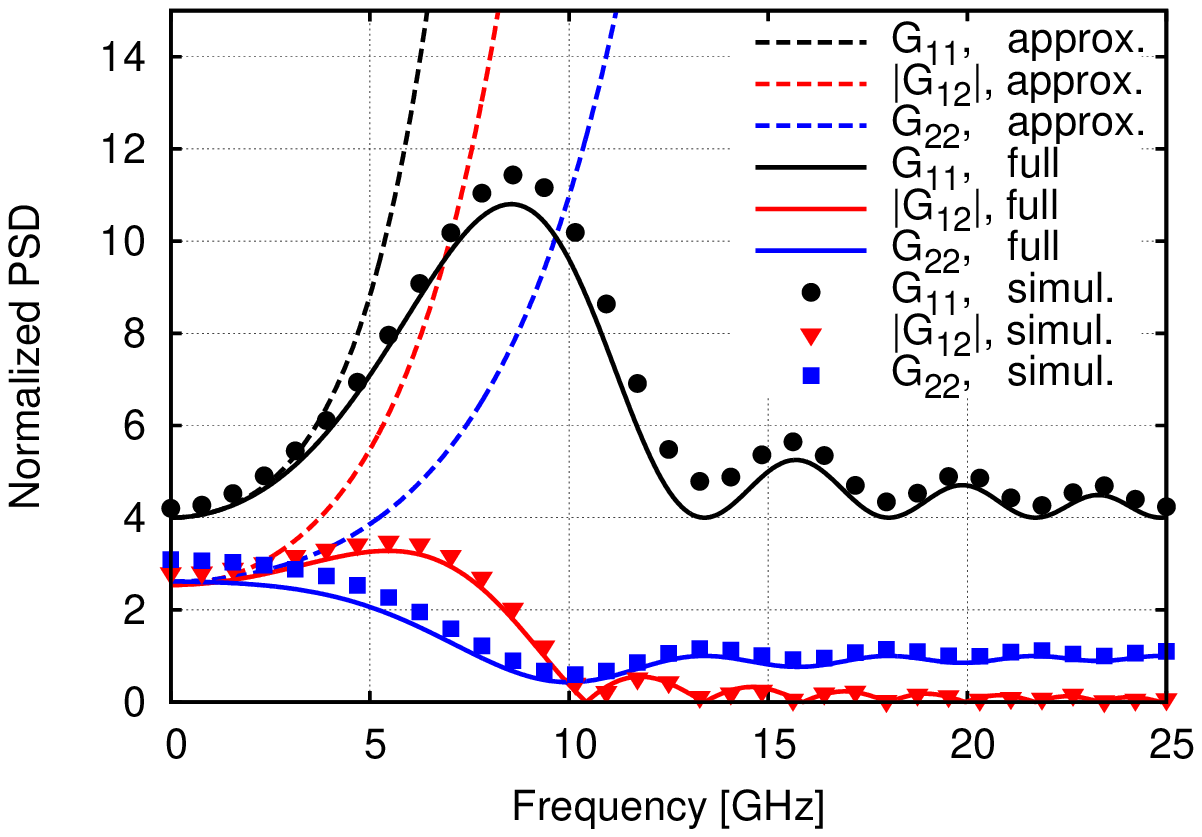}  
\caption{\label{fig:PSD_17}Normalized PSD matrix at the output of the standard fiber (anomalous dispersion $\beta_2=-21.7\,\mathrm{ps^2/km}$) for an SNR of (a) 20~dB and (b) 10~dB.}
\end{figure}

In order to investigate the impact of fiber dispersion on the accuracy of the model, the same analysis is performed also for a standard fiber with a GVD parameter $\beta_2=-21.7\,\mathrm{ps^2/km}$ (anomalous dispersion),
leaving all the other parameters unchanged. Figures~\ref{fig:PSD_17}(a) and \ref{fig:PSD_17}(b) show the corresponding results for an SNR of 20 and 10~dB, respectively.

In this case, the anomalous dispersion of the standard fiber is responsible for a significant amplification of the PSD of the power variable at some frequencies due to the presence of hyperbolic functions in Eq.\ \eqref{Tmatrix}. This corresponds to an overall increase of the noise power due to signal-noise interaction. Moreover, the higher value of the modulus of the GVD parameter confines signal-noise interaction to a narrower bandwidth. Also in this case, the accuracy of the full model is confirmed, though the discrepancy at 10~dB of SNR is slightly more relevant than in the previous case because of the aforementioned amplification of the noise power. However, the approximated model is very inaccurate because of the higher dispersion.
\par
\begin{figure}[htb]
\centering
\includegraphics[width=0.5\textwidth]{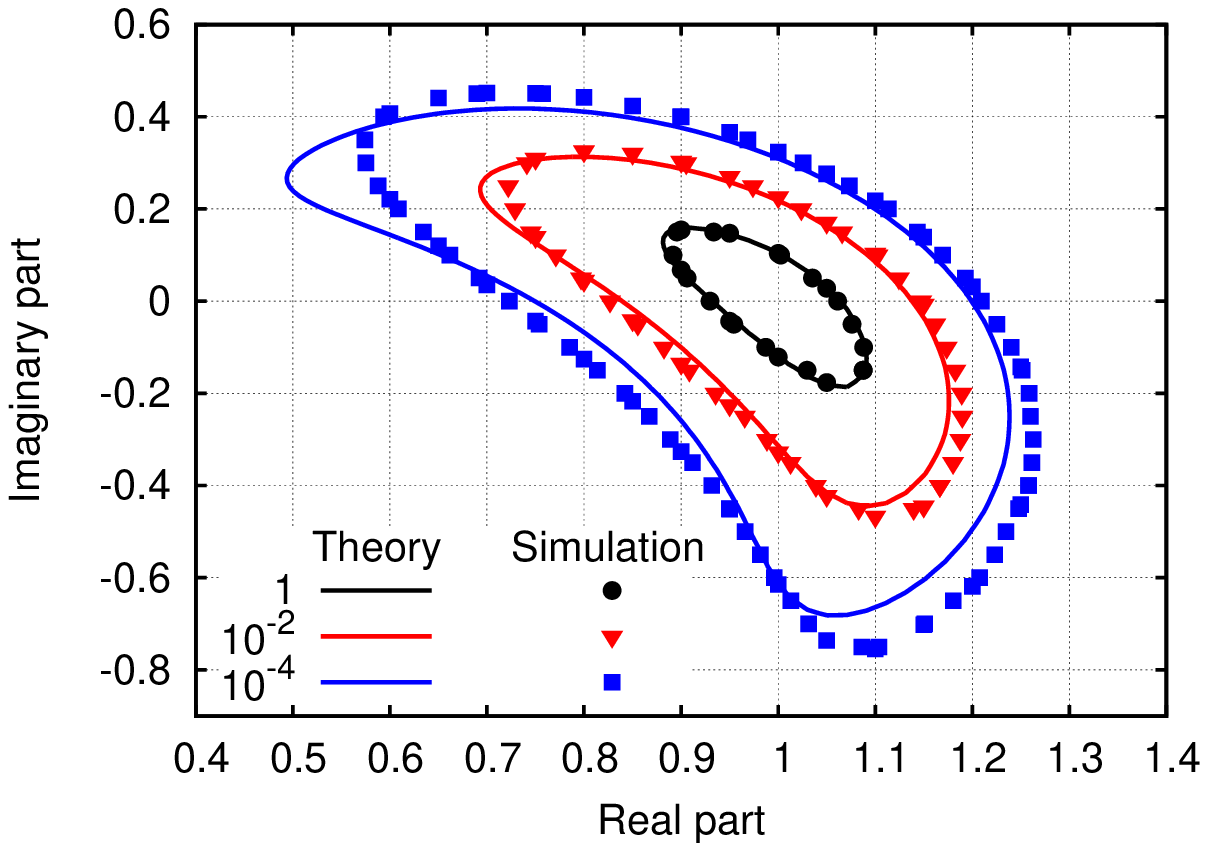}\includegraphics[width=0.5\textwidth]{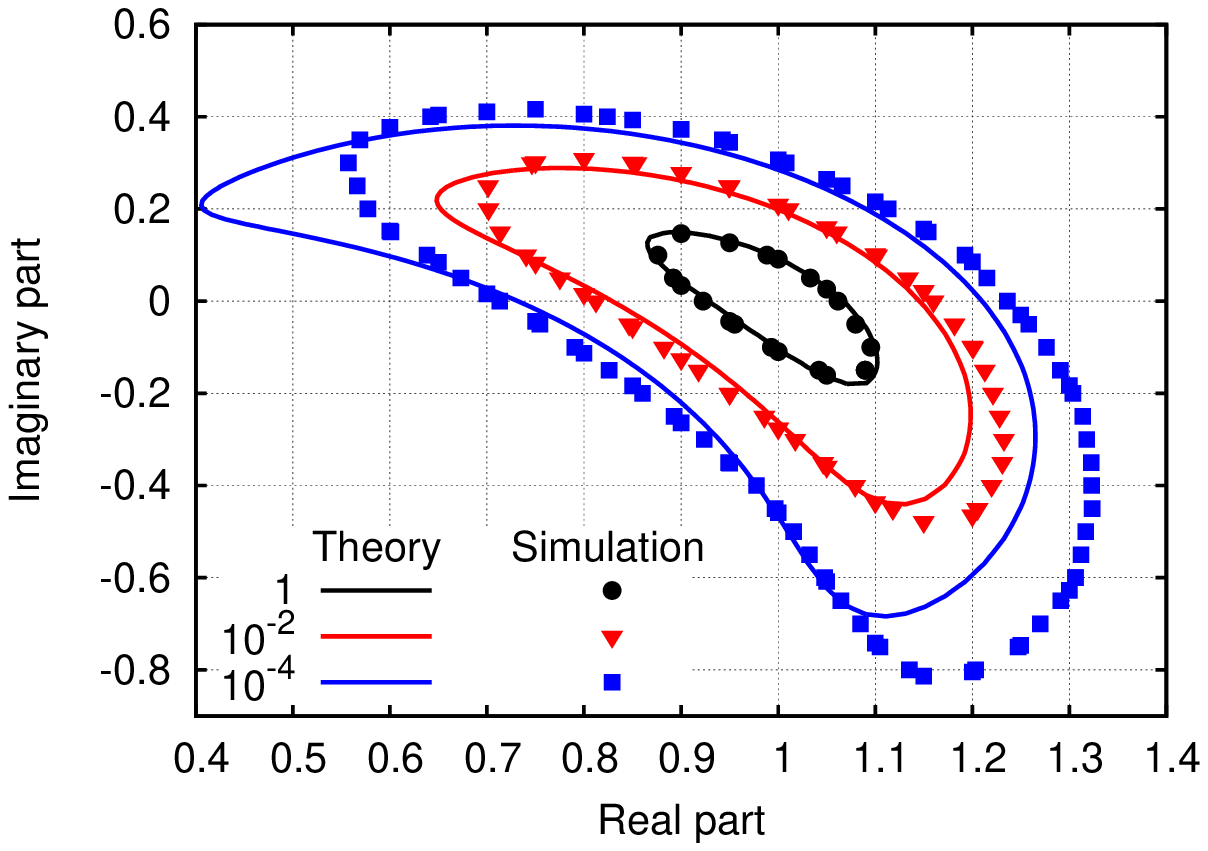} 
\caption{\label{fig:pdf}Pdf of the optical signal at the output of the (a) non-zero dispersion-shifted fiber and (b) standard fiber.}
\end{figure}

As a final test, we investigate the ability of the model to reproduce the non-Gaussian distribution of the output signal. We consider the same CW power and the same kind of fibers considered in Figs.~\ref{fig:PSD_-4} and \ref{fig:PSD_17}, an SNR of 20~dB, and reduce the noise bandwidth to 10~GHz (trying to keep all the noise within a bandwidth in which it is mostly affected by signal-noise interaction, as shown by the PSDs in the previous figures). Figures~\ref{fig:pdf}(a) and \ref{fig:pdf}(b) show the contour plots of the joint pdf of the real and imaginary parts of the (normalized and derotated) optical signal $u(z,t)e^{i\gamma\rho_0 z}/\sqrt{\rho_0}$ at the output of the fiber at an arbitrary time $t$ (as, for a CW input signal, the process is stationary) for the non-zero dispersion-shifted fiber and the standard fiber, respectively.

The solid curves are obtained according to the full first-order model, evaluating the PSD matrix of the power and phase variable as discussed before. Given the input conditions and the linearity of the propagation equations, the output variables are jointly Gaussian and their covariance matrix is evaluated by integrating the PSD matrix. The joint distribution of the real and imaginary parts is finally obtained through a change of variables.
The dotted curves are obtained by using the split-step Fourier method for the propagation of the noisy signal through the fiber, and applying the multicanonical Monte Carlo algorithm (see \cite{Gerardi:JLT2011-MMC} and references therein) to estimate the pdf with uniform accuracy down to low values.
In both cases, there is a good agreement between theory and simulations in the modal region of the pdfs, while some discrepancy appears in the tails. This is, again, due to the neglected higher-order terms, which are more relevant for higher values of the perturbation variables (i.e., in the tails of the pdf). Also in this case, the discrepancy is higher in the standard fiber because of the noise amplification that is possible with anomalous dispersion.
\section{Conclusions}
\label{S10}
We have introduced a perturbative approach, based on the hydrodynamic formulation of the NLSE, 
to the nonlinear propagation of signal and noise in an optical fiber. 
In the small-dispersion approximation, the proposed method corresponds to the semiclassical approximation of the Madelung equations in quantum mechanics. 
However, we have seen that such approximation is well-behaved only in the defocusing case (normal dispersion), for which 
the resulting first-order PDE system has a hyperbolic character. On the other hand, by removing this approximation, the method gets a more general validity and its accuracy is greatly improved.
\par
The developed formalism is rather general and, in principle, can be applied to input signals of any form.
In this paper we have considered particular input data, namely the CW signal and the solitonic pulse, 
for which simple explicit equations have been derived.
In the CW case, in particular, we have obtained explicit expressions for the PSD and pdf of the received signal 
that have been compared with numerical simulations (performed by employing the split-step Fourier method).
The numerical results confirm the soundness of the proposed approach.

\section*{Acknowledgment}
This work was supported in part by the Italian Ministry for Education University and Research (MIUR) under the FIRB project COTONE.

\end{document}